\documentstyle[11pt,newpasp,twoside,epsf]{article}

\def\vs{\vskip 5pt}
\def\ms{\vskip 2pt}

\def\eg{{e.g., }}
\def\ie{{i.e., }}
\def\etal{{et al., }}

\def\etc{{etc.}}

\def\'{^{\prime}}

\def\hq{{\hat q}}

\def\kpc{{\rm~kpc}}
\def\mpc{{\rm~Mpc}}

\def\spose#1{\hbox to 0pt{#1\hss}}
\def\lta{\mathrel{\spose{\lower 3pt\hbox{$\mathchar"218$}}
     \raise 2.0pt\hbox{$\mathchar"13C$}}}
\def\gta{\mathrel{\spose{\lower 3pt\hbox{$\mathchar"218$}}
     \raise 2.0pt\hbox{$\mathchar"13E$}}}
\def\ge{\mathrel{\spose{\lower 3pt\hbox{$-$}}
     \raise 2.0pt\hbox{$\mathchar"13E$}}}
\def\le{\mathrel{\spose{\lower 3pt\hbox{$-$}}
     \raise 2.0pt\hbox{$\mathchar"13C$}}}

\begin{document}

\title{Probing SZ Source Detection with Gasdynamical Simulations}
\author{J. Richard Bond}
\affil{Canadian Institute for Theoretical Astrophysics, University of Toronto, Toronto, Ontario, M5S 3H8, Canada}
\author{Marcelo I. Ruetalo}
\affil{Department of Astronomy and Astrophysics and CITA, University of Toronto, Toronto, Ontario, M5S 3H8, Canada}
\author{James W. Wadsley}
\affil{Department of Physics and Astronomy, McMaster University, Hamilton, Ontario, L8S 4M1, Canada}
\author{Michael D. Gladders}
\affil{Department of Astronomy and Astrophysics, University of Toronto, Toronto, Ontario, M5S 3H8, Canada}

\begin{abstract}
The huge worldwide investment in CMB experiments should make the
Sunyaev-Zeldovich (SZ) effect a key probe of the cosmic web in the
near future. For the promise to be realized, substantial development
of simulation and analysis tools to relate observation to theory is
needed. The high nonlinearity and dissipative/feedback gas physics
lead to highly non-Gaussian patterns that are much more difficult to
analyze than Gaussian primary anisotropies for which the procedures
are reasonably well developed. Historical forecasts for what CMB
experiments might see used semi-analytic tools, including large scale
map constructions, with localized and simplified pressure structures
distributed on a point process of (clustered) sources. Hydro studies
beyond individual cluster/supercluster systems were inadequate, but
now large-volume simulations with high resolution are beginning to
shift the balance. We illustrate this by applying ``Gasoline''
(parallelized Tree+SPH) computations to construct SZ maps and derive
statistical measures. We believe rapid Monte Carlo simulations using
parameterized templates centered on point processes informed by
optical and other means on the observational side, and by hydro
simulations on the theory side, should play an important role in
pipelines to analyze the new SZ field data.  We show that localized
sources should dominate upcoming SZ experiments, identify sources in
the maps under filtering and noise levels expected for these
experiments, use the RCS photometric optical survey as an example of
redshift localization, and discuss whether cosmic web patterns such as
superclusters can be enhanced when such extra source information is
supplied.
\end{abstract}

\section{Semi-analytic {\it cf.} Hydro Approaches to SZ Forecasts}

\noindent
{\bf 1.1 The Resolution of Upcoming SZ Experiments:} The Compton
upscattering of CMB photons by hot inhomogeneous (nonlinear) gas leads
to secondary CMB anisotropies $\Delta T/T (\hq ,\nu) =-2y_C (\hq
)\psi_K(h\nu /k_BT_\gamma )$ in direction $\hq$ at frequency $\nu$,
where $y_C = (\sigma_T /m_ec^2) \int n_e
k_B (T_e-T_\gamma)\, d{\tt l.o.s} $ is the Compton y-parameter and $\psi_K$ depends only upon
frequency.\footnote{Here, $\sigma_T$ is Thompson cross section, $m_e$ is the
electron mass, $T_\gamma$ is the CMB photon temperature, and $d{\tt
l.o.s}$ is the $\hq$-line-of-sight radial distance
element. $\psi_K(x)=2-(x/2)(e^x+1)/(e^x-1)$ is 1 in the Rayleigh-Jeans
region, zero at 217 GHz, negative above.} It is because $y_C$ is a
direct probe of the line integral of the electron pressure in the hot
intergalactic medium that the heavy investment in SZ experiments is
so worthwhile.

SZ observations of individual rich clusters have been possible for a
decade, are now routine (\eg Carlstrom \etal 1999), and complementary
to X-ray, optical and weak lensing observations. For resolved sources,
the surface brightness of an SZ source is independent of its
redshift. Even with the expected source evolution, this property
should make the SZ effect a valuable probe of the cluster/group
near/mid-field even at redshifts $z \sim 1$, when we expect the system
to be in a very active merging state. Thus, the era of blank field (or
ambient) SZ surveys is upon us, some targeting the $\sim 1^\prime$
resolution well-matched to the cluster/group system at $z \sim 1$,
others with $\sim 5^\prime$ resolution, probing larger sky fractions,
albeit with considerably enhanced source confusion.  Specifications of
a sample of upcoming/proposed SZ experiments are:

\ms
\noindent
{\bf Bolometer-based:} ACBAR: Viper telescope, 16 element,
multi-frequency, $\sim 4^\prime$ {\it fwhm} resolution, now;
Bolocam/CSO: 10.4m CSO telescope, 151 pixels, 150, 220, 270 GHz,
$1^\prime$, fall 2001; Bolocam on the LMT?; ACT: 3 32x32 pixel
bolometer array, $1.7^\prime$, proposed; Planck: $\sim 7^\prime$ at
150 GHz, $\sim 5^\prime$ at 220 GHz, full sky, 2007.

\ms
\noindent
{\bf HEMT-based Interferometers:} OVRO mm array: 6 dish, 10.4m, 30
GHz, now; BIMA: 10 dish, 6.1m, 30 GHz, now; CARMA: OVRO+BIMA; CBI: 13 dish,
0.9m, 30 GHz, $\sim 4^\prime$; SZA: 6 dish, 3.5m, 30 GHz (+90 GHz),
$\sim 2^\prime$; AMIBA: 19 dish, 1.2m+0.3m, 90 GHz, $\sim 2^\prime$;
AMI: 10+8 dish, 3.7m+13m, 15 GHz, $\sim 2^\prime$. The resolution can
be improved by spreading the dishes to longer baselines.

\vs
\noindent
{\bf 1.2 Historical Semi-analytic SZ Forecasting:} SZ estimates and
limits derived from them were influential ever since Sunyaev and
Zeldovich proposed the effect. First, attention was paid to
baryon-dominated (BDM) models, with entropy injection via shocks or
radiation, then to shock-heated neutrino-dominated (HDM) models, and
then the many variants of cold dark matter (CDM) models, often with
strong ``feedback'' of energy into the pregalactic or intergalactic
medium.  For example, in the eighties one of us (Bond \etal 1980s) did
HDM forecasts, first using the popular pancake treatment of structure
formation, then a better cluster-based treatment using density peaks
of various masses, including both Poisson and continuous clustering
contributions of these ``shots'' (B88). Explosion-dominated models of
structure formation in CDM, BDM models, from very massive objects,
galaxies, superconducting strings, extreme preheating, \etc \, were
also addressed in B88. All of these models became severely challenged
by the COBE/FIRAS data which gave a $10^{-4}$ 95\% CL upper bound to
the allowed fraction of the CMB radiation in a Compton cooling
distortion.

Early SZ maps using cluster/group-scale peaks in CDM models were also
constructed in the eighties (B90).  In the nineties, the calculations
of SZ maps became more sophisticated with the peak-patch technique,
which included correct spatial clustering of halos, but ``painted on''
simple parameterized pressure-profiles within the halos (Bond \etal
1990s, Bond \& Myers 1996, Bond \& Crittenden 2001 [BC01]). 

\vs
\noindent
{\bf 1.3 Hydro Approaches:} Early single-cluster SPH calculations
(Bond \etal 1990s) became more sophisticated as the codes and
computing power improved, as described for example in the ITP
``adiabatic'' cluster comparison test (Frenk \etal 2000), for which
SZ, X and weak lensing maps are shown in BC01.  Ideas of the cosmic
web interconnections of clusters of peak-patches were used to create
optimally-designed rare supercluster (TreePM-SPH) simulations (Bond
\etal 1998, reviewed in BC01) that included a proper tidal field
acting on its 104 Mpc high resolution patch (comoving lattice spacing
$1.0\mpc$, best $z$=0 resolution $20\kpc$) and cooling (but no
feedback) to see which, if any, of the X, SZ or weak lensing probes
could be sensitive to mid-field and far-field structures around
clusters and groups (\eg the far-field filamentary bridges). As
expected, we found weak lensing probes the far-field better than SZ,
which does better than X.  However, a critical issue is how to reveal
such extended-source patterns, given the projected contributions from
the clusters and environs behind and ahead of the supercluster targets
(Sec.~3).

In recent years, simulated SZ maps generated from hydrodynamical
cosmological simulations have been used to predict what the
experiments should see, mostly with emphasis on low order statistics
such as the angular power spectrum. Even with the great improvements
in computing power and codes we have seen, only a relatively small
number of hydro realizations per SZ map-making exercise are being
done, and so all such work remains statistically
incorrect. Nonetheless, we believe these approximate treatments are
useful steps along the path to that brave day when the full redshift
range relevant to the projected maps from, say, 0 to 2, is do-able,
the space tiled by contiguous simulation patches self-consistently
constructed to have coherent long-waves joining them, all at the
required resolution. This is routinely and rapidly done with the
painted-halo peak-patch approach (Bond \& Myers 1996).

\section{Approximate SZ Map-Making using High Resolution Hydro}

\noindent
{\bf 2.1 Hydro Simulations with Gasoline:} We are applying the very
efficient parallelized ``Gasoline'' code, developed by one of us (JW)
in collaboration with Joachim Stadel and Tom Quinn, to a new round of
huge LSS simulations targeting the SZ effect. This tree+SPH code uses
the smooth particle hydrodynamics method with a pure tree-based
gravity solver and has spatially-adaptive time stepping. It has been
parallelized using the MPI architecture and shows excellent
scalability. The results presented here are based on the analysis of a
200 Mpc high resolution simulation of a ``now standard'' $\Lambda$CDM
cosmology ($\Omega_\Lambda=0.7$, $\Omega_m=0.3$, $\Omega_bh^2=0.020$,
$h=0.7$, $n_s=1$, $\sigma_8$=0.90) using a ``workhorse'' level of
$256^3$ dark matter and $256^3$ gas particles (lattice spacing
$0.78\mpc$, best resolution $15\kpc$). We are now analyzing a larger
(400 Mpc box) $\Lambda$CDM simulation using $512^3$ dark matter and
$512^3$ gas particles (now with $\Omega_bh^2=0.022$), performed on a
large-memory 114 (667 MHz) processor COMPAQ SC cluster at McMaster,
that required about 80 GB of memory and took $\sim 40$ days of wall
time to run --- a ``state-of-the-art'' simulation for SZ studies.  The
$256^3$ run required 15 GB of memory and less than 6 days on 88
processors.  The specific calculations were ``adiabatic'', in the
sense that only shocks could inject entropy into the medium. Other
sources of entropy injection, from galaxy winds for example, are
expected and generally may be highly inhomogeneous, a complication
important to explore but which we ignore for this exercise.

\vs
\noindent
{\bf 2.2 SZ Maps with Experimental Beams and Noise:} Given only one
single medium-sized simulation volume whose output is densely sampled
in redshift, we create a ``pseudo-realization'' of the cosmic
structure by stacking randomly translated and rotated copies of the
same periodic box, an approach taken by other authors (e.g. da Silva
\etal 2000; Springel \etal 2001). The SZ maps are generated by
projecting the gas pressure in each copy of the box along the line of
sight and later adding all the projections. This is inadequate because
objects at various times in their history are seen again and again,
but one hopes that it is good enough for statistical exploration if
the size of the volume is not too small. As computing power improves,
the number of boxes to be brought into the mix will be larger and
larger, as will the size of the individual ones. 

\begin{figure}
\epsfxsize=11cm
\begin{center}
\epsffile{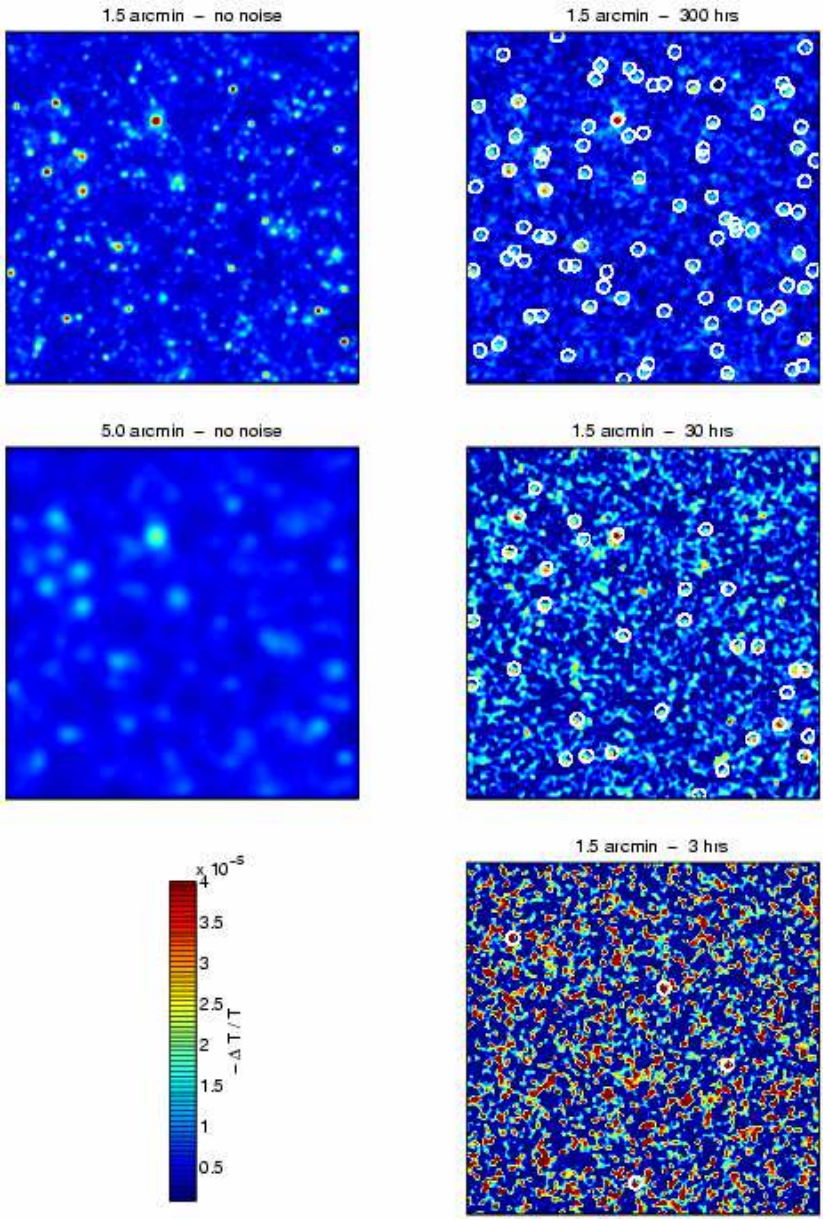}
\end{center}
\caption{The left panels show a typical $2^{\circ} \times 2^{\circ}$ SZ map generated with the
pseudo-realization method of Sec.~2.2 applied to a ``workhorse'' 200
Mpc Tree+SPH $\Lambda$CDM simulation, with 256$^3$ N-body plus 256$^3$
SPH particles, rather than the 400 Mpc $512^3+512^3$ simulation. The
maps are smoothed with {\it fwhm} filters of $1.5^\prime$, near-optimal
for the high redshift cluster/group system, and the
source-confused $5^\prime$ appropriate to ACBAR, Planck, \etc\,  The right panels
show the same $1.5^\prime$ map after an increasing amount of
instrumental noise has been applied - mimicking proposed AMIBA
specifications and integration times from ``deep'' (300 hrs) to ``shallow''
(3 hrs). The 30 hr AMIBA noise level roughly corresponds to $\sim 200$ hrs
for current Bolocam specifications. Circled objects were found in the
maps using an efficient algorithm developed to identify optical
galaxies in noisy CCD images (Yee 1991). If CMB primary and Galactic
foreground signals are added, better designed ``optimal''
filters are needed (\eg BC01). Note that only a few objects can be
identified by the algorithm in the ``shallow'' map. Knowing source
positions from non-CMB surveys can improve identification
considerably. }
\label{fig:SZ_MapsSmoothingNoise}
\end{figure}

Fig.~\ref{fig:SZ_MapsSmoothingNoise} shows the effect of smoothing for
beams spanning the range of the proposed SZ experiments, and also how
the addition of varying levels of homogeneous noise obscure the SZ
source structures. The maps were made at Rayleigh-Jeans wavelengths,
so $\Delta T/T \approx -2y_C$. Each experiment will have its own special
$\ell$-space filtering which truncates at low $\ell$ as well as high,
but for our purposes this simple Gaussian-filtered map will do to make
our points. Note the high source confusion expected for experiments
with beam sizes $\gta 5^\prime$.

\vs
\noindent
{\bf 2.3 Finding SZ Sources Using SZ Maps Alone:} The preliminary
approach to source identification shown in
Fig.~\ref{fig:SZ_MapsSmoothingNoise} used an algorithm borrowed from
optical astronomy developed for galaxy identification by Yee (1991),
but more optimal object identification algorithms appropriate for
separating the SZ signal from primary CMB signal and Galactic
foregrounds as well as noise (following the ``optimal'' algorithms
described in BC01) are needed to analyze scanned maps, both
single-dish bolometer-style and interferometer-style using
mosaicing/drifting.
 
\begin{figure}
\epsfxsize=8cm
\vspace{-0.3in}
\begin{center}
\epsffile{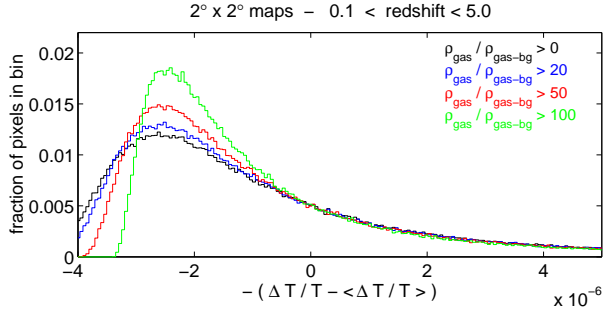}
\end{center}
\vspace{-0.3in}
\caption{Histograms of the fraction of pixels in a series of SZ maps
corresponding to density cuts $1+\delta_{g,cut}$ $\in \{0, 20, 50, 100\}$.
The subtracted means
$2 \overline{y_C}$ are $\{5.80,4.91,4.15,3.46\}\times 10^{-6}$ respectively. 
Convergence beyond the means
to a nearly common histogram suggests that a localized pressure-source
model should capture much of the observable range. How many of the
far-field pixels (where the histograms deviate) should be added to
adequately extend the sources depends upon the noise and filtering
characteristics of each experiment. }
\label{fig:SZ_MapsHistogram}
\end{figure}

\vs
\noindent
{\bf 2.4 Where Does Most of the Contribution Come From?}  It is clear
visually that most of the contribution to the noise-free SZ maps comes
from concentrated sources. To address this quantitatively, we applied
overdensity cuts $\delta_{g,cut}$ to the gas particles in our SPH
simulations that were allowed to contribute to the SZ effect, and
compared the resultant maps visually and using the one-point
distribution function of the pixel values of the temperature
fluctuations (Fig.~\ref{fig:SZ_MapsHistogram}). Except for the density
cut, all other aspects of the ``stacking of boxes'' map-making method
were treated in the way outlined above to facilitate comparison.
Since SZ experiments typically filter long wavelengths (some much more
than others), subtracting off the means $\overline{\Delta T}$ of the
maps from the pixel values ${\Delta T}_p$ is a good way to compare
what the effect of the cuts is. The histogram sequence with varying
$\delta_{g,cut}$ in Fig.~\ref{fig:SZ_MapsHistogram} shows that for
$y_C$ above $\overline{y_C}$ convergence occurs for
$\delta_{g,cut}\lta 100$. Our $\delta_{g,cut}$ SZ maps show that the
pixels just below this convergence point, where the deviations start
to appear, are contiguous to the high pixels, and define the
mid-field/far-field skirt that would define the pressure profile for
the extended-source model. How far one needs to go is very dependent
upon the noise and filter characteristics of the various experiments
that we are exploring. At lower levels, the pixels break out into the
distributed filamentary gas (\ie outside of groups.) SZ maps of thin
slabs in redshift (with density cuts) are proof of this
extended-object-contribution picture.  Low mass groups (random
superpositions) populate the interfilamentary regions in deep SZ maps,
although they would be filtered out by experiments and not be
discernible from the interfilamentary gas .

\begin{figure}
\plotone{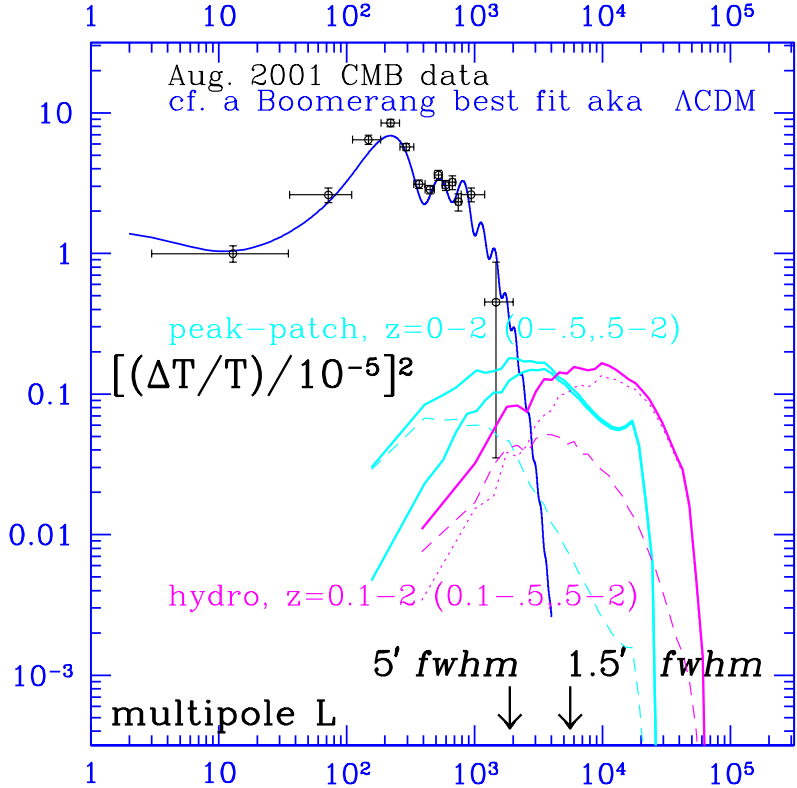}
\caption{SZ power spectrum ${\cal C}_\ell$ for one of our 4 deg$^2$ SZ
maps generated from our $200$ Mpc $\Lambda$CDM simulation. It is
compared to a (scaled) ${\cal C}_\ell$ for a 25 deg$^2$ peak-patch
simulation using a slightly different $\Lambda$CDM cosmology. At
$5^\prime$ resolution, $z < 0.5$ sources dominate, but at
$1^\prime$-$2^\prime$ resolution, the high $z$ cluster/group system
dominates. Sample variance affects the hydrodynamic result at lower
$\ell$, and at high $\ell$, short distance structure is treated
differently, reflecting computational and feedback/cooling physical
issues still unresolved in the subject. Just as in
Fig.~\ref{fig:SZ_MapsHistogram}, the range included all $z \le 2$
contributions to the Tree+SPH computation, but the adiabatic
calculation ensures that big galaxies (at the resolution edge) and
very small groups contribute, especially at higher $\ell$, probably
more than they should. The peak-patch groups had a mass cut imposed, a
sharp-threshold model of group winds motivated by this uncertain
physics: hence the smaller ${\cal C}_\ell$ at high $\ell$ is
indicative of what energy injection might do. The lower $\ell$ power
is sensitive to whether nearby clusters are in the map or not, more
likely in the 25 deg$^2$ case than in the 4 deg$^2$ case because of
smaller sample variance. As well, all large-scale tidal power is
included in the peak-patch algorithm, whereas the Tree+SPH periodicity
results in power truncation at the fundamental mode of the box.}
\label{fig:SZ_PowerSpectra}
\end{figure}

\vs
\noindent {\bf 2.5 Thermal SZ Power Spectra:}
Fig.~\ref{fig:SZ_PowerSpectra} contrasts the current CMB data on
primary anisotropies with the SZ ${\cal C}_\ell$ from our $\Lambda$CDM
SZ maps, both the small ones using the ``stacking of boxes'' and the
larger scale ones using peak-patches (BC01).  Even though power is
concentrated in sources, the ${\cal C}_\ell$ spectrum derived from our
SZ maps does show the angular scales we would most like to probe. In
Fig.~\ref{fig:SZ_PowerSpectra} we can see that at Planck resolution,
$\ell_s \sim 1600$, $z < 0.5$ sources are the most prominant
contributors, but at the $1^\prime$-$2^\prime$ scale of ground-based
bolometer arrays and interferometers, the $z \gta 0.5$ cluster system
dominates. At high $\ell$, the fate of gas in small clusters and
groups is an ongoing uncertainty, reflected in the location of the
${\cal C}_\ell$ peak and the $\ell$ behaviour beyond. The overall
${\cal C}_\ell$ magnitude is quite sensitive to cosmological
parameters, in particular $\sigma_8$ and its evolution.

\section{Simulating Source Localization and Removal in SZ Maps}

\noindent {\bf 3.1 The Needle in a Haystack Problem:} To address
whether physically-connected supercluster concentrations can be
identified in the pressure-projected SZ maps, we generated individual
SZ maps for the largest superclusters found at intermediate redshifts
in the simulated box (\eg the largest at $z$=0.82), and superimposed
them on top of our pseudo-realization SZ maps. We found that the
strong, highly concentrated, sources still stand out, but not the
filamentary structures: in fact, many apparent filamentary sources
that do stand out are actually due to random superposition.

\vs 
\noindent {\bf 3.2 Templates from non-CMB Surveys to Localize in
Redshift:} As well as ``cleaning'' future SZ maps of other CMB signals
and noise, we hope it will be possible to remove the contribution from
foreground and background concentrated SZ sources to allow access to
structures in a target redshift range. This will not be possible
without adding extra information, \eg from deep surveys in the
optical, X-ray and weak-lensing. The mass, redshift, size, orientation
and other information on optically-detected clusters can be used, \eg
to place parameterized intra-cluster gas profiles at their
locations. Although this would only allow a (rather dirty) form of
cleaning to enhance SZ structures, in conjunction with Monte Carlo
error estimations (and follow-up observations), with appropriate
algorithm development it would be a dramatic improvement over what can
be done with field SZ maps alone.

We are studying the viability of this approach by coupling our
simulated SZ maps to the cluster detection characteristics of the very
successful 100 deg$^2$ (22 patches X 5 deg$^2$) optical photometric
``Red-Sequence Cluster Survey'' just completed by Yee and Gladders
(2001).  Proposed optical surveys such as RCS2 or VISTA, covering 1
and 2 orders of magnitude larger areas, are planned for the next five
years.

\begin{figure}
\plotone{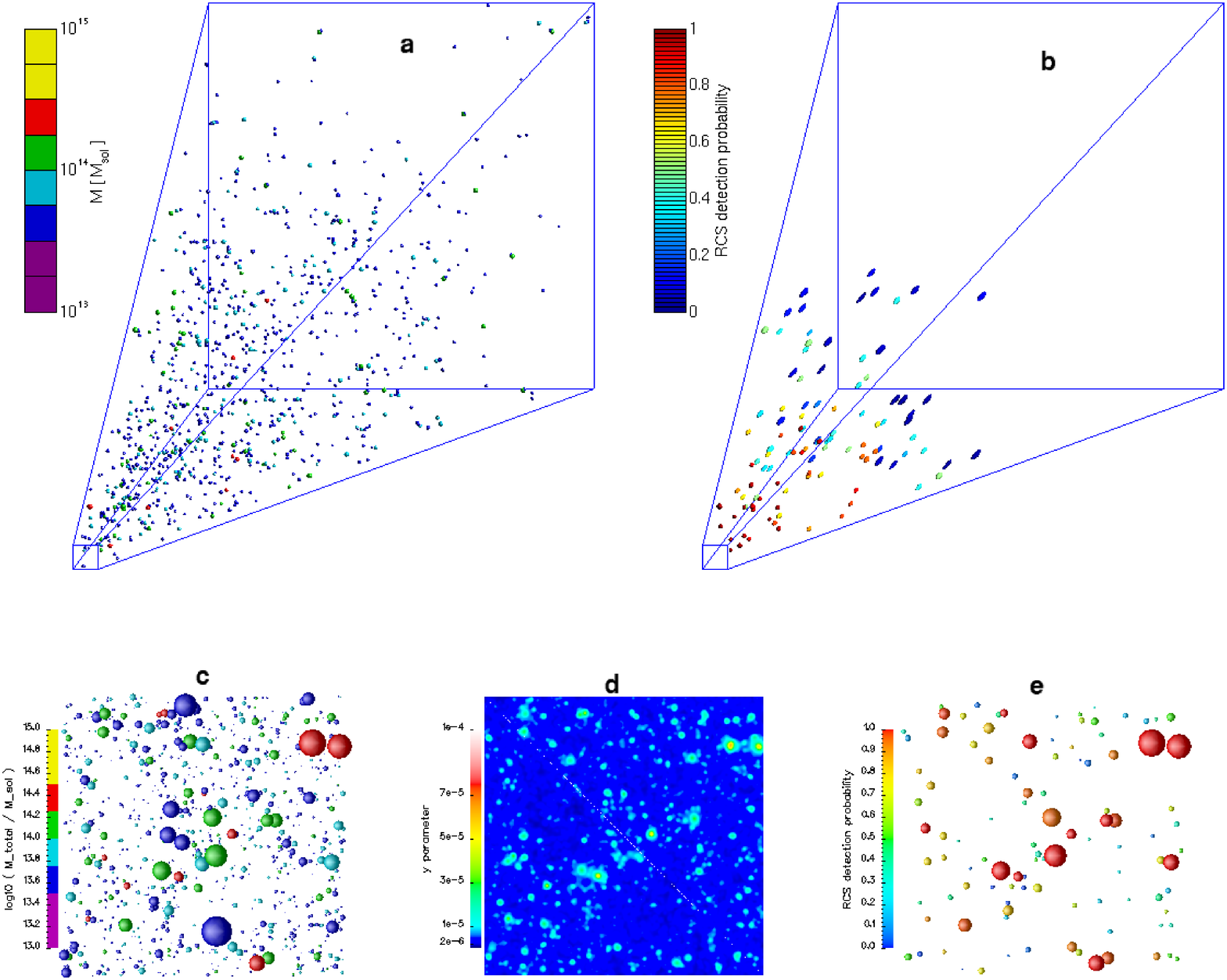}
\caption{The catalogue of clusters up to $z=2.5$ in a $2^{\circ} \times 2^{\circ}$ pyramid
pseudo-realization generated from our $200$ Mpc $\Lambda$CDM
simulation. The radial direction is comoving position. In (a),
colour and size visualize cluster/group mass and size found with a 
halo identification algorithm. In (b), colour
visualizes the optical detection probabilities of the clusters  and
length visualizes the uncertainty in redshift localization, using
parameters appropriate to the 100 deg$^2$ RCS survey of Yee and
Gladders (2001). Both decrease with
mass. For redshifts beyond unity the uncertainty in redshift is about
the size of the box of our hydrodynamical simulations, 200 Mpc, with
of course much better angular positioning. (c) and (e), the
projections of (a) and (b) onto 2D maps should be contrasted with the
$1.5^\prime$-smoothed SZ map of the same pseudo-realization:
source-blending due to superpositions is evident. For clarity, only clusters of $M > 10^{13.5}$ M$_{\sun}$ are shown in (a) and (c) and only clusters of non-zero detection probability are shown in (b) and (e). In spite of the RCS
uncertainties, it is clear that such optical surveys can provide
powerful templates for source-finding, source-deconvolution, and
redshift localization. }
\label{fig:LSSPyramidPlusRCS}
\end{figure}

When a pseudo-realization (pyramid) of the cosmic structure is
generated by stacking copies of the simulation box, we not only create
the corresponding simulated SZ maps, but also generate a catalogue of
the clusters/groups in the realization by properly translating,
rotating, \etc\, the halos identified in the box with a group-finder
at each redshift. Fig.~\ref{fig:LSSPyramidPlusRCS} shows how the
clusters in one such pseudo-realization are related to the
corresponding SZ map and a visualization of the detectability and
localization of the clusters if we had a photometric survey with the
RCS characteristics.

\vs
\noindent {\bf 3.3 SZ Maps Using Cluster Catalogues:} Since the
dominant contribution to SZ maps is due to concentrated sources, as
shown above, one can also create fast Monte Carlo simulations by using
catalogues of halos identified in very large volume (but low
resolution) N-body only simulations or generated by the ``peak-patch''
technique (Bond \& Myers 1996). The intracluster gas is modeled by
``painting'' simple parameterized profiles on the halos, though with
the improvement in progress of using (variously averaged) pressure
profiles of the clusters identified in our high resolution
boxes. These approaches offer the advantage of allowing a much faster
generation of cosmological realizations of the structures most
relevant for the study of the SZ effect. Moreover, rare events in the
cosmic structure, missing in periodic box hydro simulations, are now
accurately included because coherent box-to-box long waves are. The
drawback is, of course, that much hinges on profile choice.

\vs
\noindent {\bf 3.4 Summary and Future Explorations:} We described here
our use of high resolution hydro simulations to develop the ``extended
source'' model of the highly non-Gaussian SZ distribution. We plan to
generate more and higher resolution hydro simulations with entropy
feedback and cooling included to refine our SZ predictions. For the
adiabatic calculations shown here, we have demonstrated that, although
gas distributed in filaments and membranes in the IGM have
non-negligible SZ signals, the observable range for planned
experiments is dominated by the ``extended-source'' picture. We also
addressed the issue of source identification in SZ maps of the size,
resolution and noise level that mimic planned observing campaigns, and
explored how optical photometric surveys such as the ``Red-Sequence
Cluster Survey'' can be used to help to find SZ sources and localize
them in redshift. This could allow us to identify supercluster
structures, \ie to deproject SZ maps. Indeed, given that large area
optical surveys will be available, an argument can be made for
catalogue-targeted rather than blank-field SZ observations. We expect
both strategies will be used.

To understand the complex residual maps after unwanted extended-source
removals, a large number of fast Monte Carlos will be essential,
precluding a hydro-only approach with current computational power.
The extended-source approach applying constrained mean
pressure-profiles to halos identified in ultra-large low-resolution
N-body simulations or by the peak-patch technique, calibrated with the
high resolution hydro results, is a worthwhile path to further
explore. The snapshot of the work described here mostly shows that
there is much to develop in SZ numerical simulations and in analysis
tools to directly confront the exciting new SZ datasets coming down
the pipe.

\end{document}